\documentclass[review,sort&compress]{elsarticle}

\usepackage{amssymb}
\usepackage{epsfig}
\usepackage{graphics}
\usepackage{graphicx}
\usepackage{subfigure}
\usepackage{exscale}
\usepackage{latexsym}
\usepackage{makeidx}
\usepackage[section]{placeins}
\usepackage{xtab}
\usepackage{epsf}
\usepackage{bbm}

\usepackage{lineno}

\journal{Nuclear Instruments and Methods in Physics Research Section A}

\begin{document}
\begin{frontmatter}

\title{Upgrade of compact neutron spectrometer for high flux environments}

\author{M.~Osipenko$^a$,
A.~Bellucci$^d$,
V.~Ceriale$^b$,
D.~Corsini$^b$,
G.~Gariano$^a$,
F.~Gatti$^b$,
M.~Girolami$^d$,
S.~Minutoli$^a$,
F.~Panza$^a$,
M.~Pillon$^c$,
M.~Ripani$^{a,f}$,
D.M.~Trucchi$^d$,
}

\address{
$^a$ \it\small INFN, sezione di Genova, 16146 Genova, Italy, \\
$^b$ \it\small Dipartimento di Fisica dell'Universit\`a di Genova, 16146 Genova, Italy, \\
$^c$ \it\small ENEA, Frascati, 00044 Italy, \\
$^d$ \it\small CNR-ISM, Monterotondo Scalo, 00015 Italy, \\
$^f$ \it\small Centro Fermi, Roma, 00184 Italy. \\
}

\begin{abstract}
In this paper a new version of $^6$Li-based neutron spectrometer for high flux environments is described.
The new spectrometer was built with commercial single crystal Chemical Vapour Deposition diamonds of electronic grade.
These crystals feature better charge collection as well as higher radiation hardness.
Ohmic metal contacts were deposited on the diamonds suppressing build-up of space
charge observed in the previous prototypes.
New passive preamplification of signal at detector side was implemented to improve the resolution.
This preamplification is based on RF transformer not sensitive to high neutron flux.
Compact mechanical design allowed to reduce detector size to a tube
of 1 cm diameter and 13 cm long.
The spectrometer was tested in thermal column of TRIGA reactor and at DD neutron generator.
The test results indicate an energy resolution of 72 keV (RMS)
and coincidence timing resolution of 68 ps (RMS).
The measured data are in agreement with Geant4 simulations except for larger energy loss tail
presumably related to imperfections of metal contacts and glue expansion.
\end{abstract}

\begin{keyword}
neutron spectrometer \sep diamond detector \sep fission spectrum

\PACS 29.30.Hs \sep 29.40.Wk

\end{keyword}

\end{frontmatter}

\section{Introduction}\label{sec:intro}


Neutron spectroscopy in high flux environments such as fission or fusion reactors is a very challenging task.
The standard methods make use of fission chambers or activation foils.
However, both of these methods are indirect and they are subject to large systematic uncertainties.
They are indirect because observable quantities are related to integrals
of neutron spectrum and the unfolding of these integrals gives rise to uncertainties.
The spectroscopic application of conventional gas-filled proportional counters with various
converters, like $^3$He, CH$_4$ and BF$_3$, is limited to very low neutron energies,
where the range of produced charged particles in gas still lies within the detector volume.
Scintillators and standard semiconductor based detectors suffer from a strong radiation damage.
Moreover, scintillators feature strong quenching for low energy ion recoils affecting
the energy reconstruction. Diamond is the most radiation hard semiconductor
offering a number of beneficial properties~\cite{cvd_rad_hard}.

In Ref.~\cite{sdw_calib} a new neutron spectrometer for such measurements was proposed.
It was based on a sandwich of two diamond sensors enclosing $^6$Li converter.
Such device allows to measure neutron energy directly on event-by-event basis
applying the energy conservation law. The first prototype of the spectrometer
was calibrated in Ref.~\cite{sdw_calib} at two neutron energies and tested in fast fission reactor in Ref.~\cite{sdw_tapiro}.
These experiments revealed a series of issues leading to a degradation of spectrometer performances.
In particular, the fast build-up of space charge limited charge collection stability
of the spectrometer to relatively low neutron fluences $<10^{10}$ n/cm$^2$.
Selected diamonds were not very radiation hard and in fact after
the experiments, corresponding to accumulated fluence of fast neutrons about $10^{14}$ n/cm$^2$,
the spectrometer showed an increased dark counting rate.
The energy resolution was limited by few meters long cables between the spectrometer and its first amplifier.

All these issues were dealt with in the work described in this article, resulting in
development of a more advanced spectrometer prototype.
To suppress space charge build-up ohmic contacts were deposited.
Selection of higher quality diamond crystals also affected charge collection
as well as the radiation hardness of the spectrometer as explained in Ref.~\cite{sdw_dt}.
Implementation of passive amplification scheme near the sensor
allowed to improve signal-to-noise ratio and therefore resolution
in spite of long cables and fast electronics used.
Furthermore, resolution on the coincidence time between two diamond
sensors was improved by more than one order of magnitude.
In the following sections the new spectrometer is described in details
along with new characterization measurements performed
at TRIGA reactor with thermal neutrons
and at Frascati Neutron Generator (FNG) neutron source with 2.5 MeV neutrons.

\section{Detector Upgrade}\label{sec:det}
In the new prototype of sandwich spectrometer, commercial, electronic grade
single crystal CVD diamonds from E6~\cite{cvd_e6} were used.
The diamonds were 300 $\mu$m thick and had surface area of $3\times 3$ mm$^2$.
Almost the entire top and bottom surfaces were covered with thin metal contacts.
All the samples were cleaned in a strongly oxidizing solution
(H$_2$SO$_4$ : HClO$_4$ : HNO$_3$ in the 1:1:1 ratio, 15 minutes at boiling point),
followed by rinsing in aqua regia (HCl : HNO$_3$ in the 3:1 ratio, 5 minutes at boiling point)
and ultrasound sonication, in order to remove organic and metallic impurities,
possible non-diamond contents, and residual debris.
Metallization procedure on top and bottom surfaces of diamond samples
consisted of the formation of a 3 nm-thick Diamond-Like Carbon (DLC) layer by
energetic (700 eV) Ar+ ion bombardment, able to induce amorphization of the diamond surface.
DLC ultra-thin layers at diamond-metal interface were indeed demonstrated
to improve contact ohmicity and stability under high-flux irradiation~\cite{Trucchi1}.
Subsequently, a 100 nm-thick Au layer was grown in situ
by RF magnetron sputtering (RF power 200 W, base pressure $10^{-6}$ mbar, Ar$^+$ pressure $2.3\times 10^{-2}$ mbar).
Lateral dimensions of the contacts (2.8 x 2.8 mm$^2$) were defined by stainless steel
shadow masks positioned on the diamond surface during both the DLC layer formation
and Au layer deposition processes.
The improved ohmicity of the electric contacts, mostly induced by the quality of
the DLC layer produced on the diamond surface, allowed for the fabrication
of ionizing radiation detectors~\cite{Trucchi2,Trucchi3,Trucchi4}
with reduced build-up of space charge under the device electrodes.
At the borders of contact area two 200 $\mu$m wide and 150 nm thick
strips were added, similar to those in Ref.~\cite{sdw_calib}.

On one diamond 100 nm thick LiF film enriched with $^6$Li to 96\% was thermally evaporated on the metal contact.
The evaporation was performed inside an evaporation chamber
evacuated down to a pressure of 10$^{-6}$ mbar. The LiF powder was poured
inside a tungsten crucible, electrically connected to its power supply.
The samples were mounted on a sample holder located over the crucible
with a quarz microbalance being placed on the same plane. Margins of diamond surface were covered
with the same stainless steel mask used for metal contact deposition.
The thickness was controlled
during deposition by the microbalance. The expected ratio of the subtended
solid angles from the sample holder and the microbalance was estimated to be about 0.4.
After the deposition the effective thickness of deposited film was measured to be 100 nm
using an interferometer microscope.

\begin{figure}[h]
\begin{center}
\includegraphics[bb=1cm 0cm 20cm 26cm, angle=270, scale=0.3]{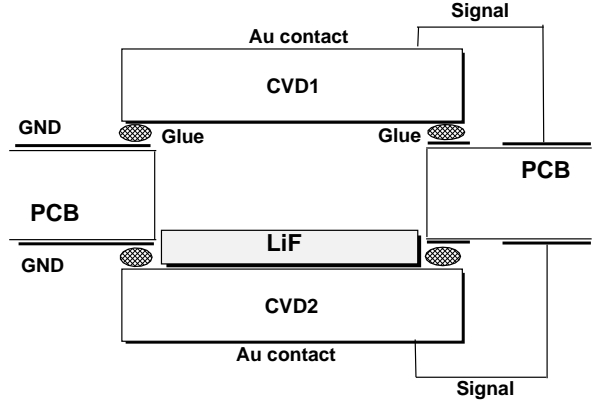}
\caption{\label{fig:sdw_geom}Vertical projection of the spectrometer drawing (not in scale) around
its sensitive part. CVD1 and CVD2 indicate two diamond crystals, each of them have metal contacts
on its top and bottom surfaces. Diamonds are glued to PCB at the edges of square hole
leaving large fraction of detector area open.}
\end{center}
\end{figure}

The diamond sensors were glued with conductive glue E-solder 3025~\cite{esolder} at opposite sides
of a 250 $\mu$m thick double-face PCB, above a square through-hole of size
slightly smaller than the diamond dimension,
as shown in Fig.~\ref{fig:sdw_geom}.
This procedure requested development of a special tool for holding the diamonds
at their expected positions above the hole during curing time.
This way most of diamond inner (w.r.t. PCB inserted in the middle)
surfaces were not obscured
by PCB and charged particles could travel from one diamond to the other, losing energy only in about 300 $\mu$m of air.
On the PCB circuit the inner diamond contacts were set to ground by conductive glue, while the outer contacts
were connected to high voltage bias and signal readout vias by wedge bonding.

\begin{figure}[h]
\begin{center}
\includegraphics[bb=1cm 0cm 18cm 26cm, scale=0.3, angle=270]{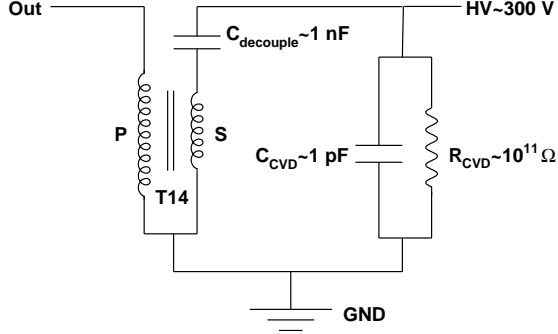}
\caption{\label{fig:sdw_preamp}Electrical scheme of diamond detector readout.
Diamond detector is modeled by the resistance $R_{CVD}$ and capacity $C_{CVD}$ connected in parallel.
Current signal from the detector is transformed into voltage in the transformer T14.}
\end{center}
\end{figure}

Electronic grade single crystal CVD diamond features a very large resistance,
of the order of $10^{10}\div10^{11}$ $\Omega$. Thus it can be approximated
as an ideal current source with infinite intrinsic impedance.
In the present application the Si-based amplifier cannot be installed
near to the detector due to large flux of fast neutrons. These neutrons
would damage Si-based electronics much faster than the diamond sensor.
Few meter long coaxial cables must therefore be used to carry detector signals
outside of high irradiation region. Given the typical impedance of coaxial cable (50 $\Omega$ for RG174)
it is recommendable to transform the signal maximizing its voltage amplitude before entering into the cable.
This was accomplished by means of a fast RF transformer as shown in Fig.~\ref{fig:sdw_preamp}.
For this purpose we selected Mini-Circuits T14-1-KK81 RF transformer~\cite{Mini-Circuit},
which features relatively high impedance ratio of 14 and wide bandwidth of 150 MHz.
This allowed to amplify voltage amplitude of the signal by a factor 3
shifting its main frequency into the range less affected by cable attenuation.
In fact the transformation renders the output signal very similar
to Si detector response, recovering the difference in the number of eh-pairs
produced per unit energy and reducing rise-time and fall time.

PCB had dimensions of 8.5$\times$80 mm$^2$ and it was inserted into aluminum tube
with external diameter of 10 mm, which provided the shielding against EMI.
The dimension of the tube was chosen for compatibility with small channels
in fast reactors like Venus-F~\cite{SCK-CEN}.

The output signals from the PCB were connected via 3.5 m RG174 cables to Wantcom WBA0010-45A~\cite{wantcom}
amplifiers. These amplifiers provide gain of 45 dB
introducing minimal amount of noise. Also the lower edge of accepted frequency band of 10 MHz
allowed to suppress environmental noise.
The signals were amplified further by a Phillips Scientific 771 amplifier.
The final signals were acquired by SIS3305 digitizer at sampling rate of 5 Gs/s.
The custom Data AcQuisition (DAQ) system
was running on Concurrent Tech. VX813-09x single board computer
saving data to a fast SATA SSD. More details on DAQ may be found in Refs.~\cite{sdw_calib,sdw_dt}.

\section{Measurement at TRIGA reactor}\label{sec:triga}
The spectrometer was carefully calibrated in the well known thermal neutron flux at LENA of Pavia University~\cite{lena_pv}.
The detector was installed in the TRIGA reactor thermal column inside
a special, low flux cavity nearby a small, calibrated fission chamber~\cite{altieri_fc}.
In this location neutron flux reaches $10^{8}$ n/cm$^2$/s at the maximum reactor power of 250 kW.
Besides the near fission chamber, the neutron flux was cross checked
through in-core monitoring system~\cite{triga_mon} whose relation
with thermal column flux is given in Ref.~\cite{triga_flux}.

\begin{figure}[!h]
\begin{center}
\includegraphics[bb=3cm 0cm 20cm 26cm, scale=0.35, angle=270]{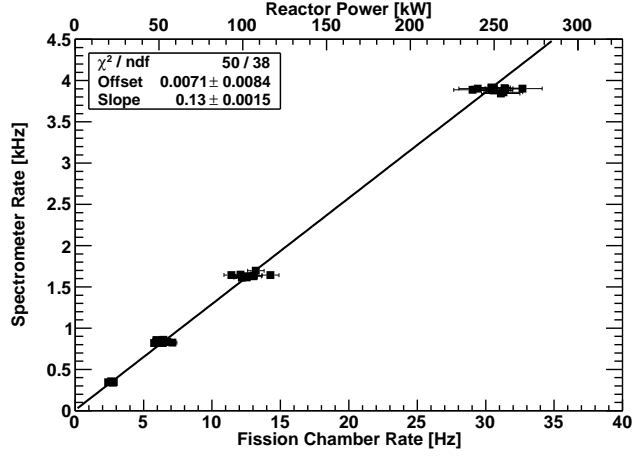}
\caption{\label{fig:pavia17_fc_swd_rates}Correlation between neutron spectrometer event rate,
near fission chamber counting rate and reactor power. Uncertainties are statistical only.}
\end{center}
\end{figure}

The data were obtained for reactor power varying from 20 kW up to 250 kW
and correlation between three detectors were studied.
The spectrometer rate linearity with neutron flux variation measured with near fission chamber
and with TRIGA power
is shown in Fig.~\ref{fig:pavia17_fc_swd_rates}.

%

\begin{figure}[!h]
\begin{center}
\includegraphics[bb=4cm 0cm 20cm 26cm, scale=0.35, angle=270]{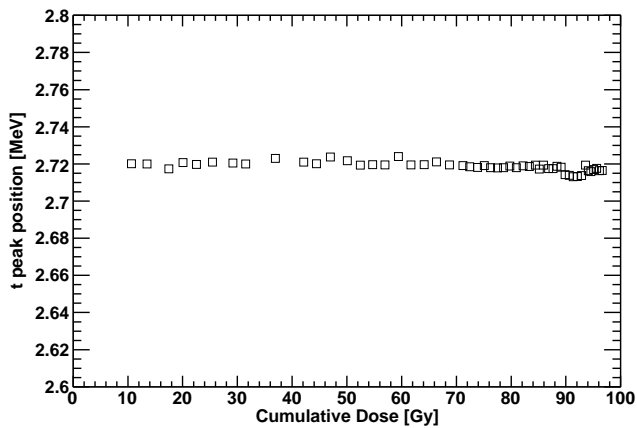}
\caption{\label{fig:et_stability}Stability of charge collection measured as the average energy
deposited by $t$ as a function of $\alpha$ and $t$ dose accumulated in the irradiated volume of spectrometer.
Uncertainties are statistical only.}
\end{center}
\end{figure}

The previous spectrometer prototypes exhibited insufficient stability of Charge Collection Efficiency (CCE)
due to rapid build-up of space charge~\cite{sdw_tapiro}. With this new prototype no visible
space charge effects were observed. CCE stability was obtained from the average energy
deposited by $t$, produced by thermal neutrons through $n(^6Li,t)\alpha$ reaction,
as a function of absorbed dose as shown in Fig.~\ref{fig:et_stability}.
For thermal neutrons in the column the dose was essentially given by $\alpha$ and $t$ particles
produced in $^6$Li and involved only a small fraction of entire diamond thickness:
3.5 and 21 $\mu$m, respectively. Thus the dose was calculated for this volume
irradiated by $\alpha$ and $t$. Small variations of the peak position, of the order of few keV,
were due to fluctuations of EMI noise during the run time altering the shape of $t$ peak.
The spectrometer was also irradiated with 4.7 kBq $^{241}$Am $\alpha$ source for 24 hours,
showing no visible change of CCE.

%

Trigger of DAQ system was set on the logical sum of the two diamond sensor discriminators.
Individual sensors had the thresholds calibrated to 1 MeV of deposited energy
as the best trade-off between electronic noise suppression and detection efficiency.
Also few runs with trigger set to a coincidence between two diamond sensors within 64 ns window
were recorded for a cross check.
Because thermal neutrons and $\gamma$ could not produce signals in diamonds above 1 MeV
threshold, only about 25\% of uncorrelated triggers had no coincidence.
Half of these events had $t$ detected with energy reduced by about 60 keV with $\alpha$ lost.
The other half had $\alpha$ detected but its energy distribution had a large tail
increasing towards the threshold indicating a high energy loss in non sensitive parts
of the spectrometer.
These events were discarded in the following analysis.

The spectrometer response to the TRIGA thermal neutron flux was modeled using Geant version 4.10.2~\cite{geant4}.
The simulated geometry was slightly simplified, in particular conductive glue and bonding wires were
not modeled. Also metal contacts on the diamond surface were assumed to be uniform.
The TRIGA thermal column flux with spectrum from Ref.~\cite{triga_flux} was generated isotropically on the spherical
surface around the spectrometer of area about 0.64 cm$^2$. The same trigger conditions
were applied to simulated events selecting those which deposited in active detector volume
an energy above 1 MeV threshold.
The simulations were normalized to the neutron fluence accumulated by fission chamber
nearby the spectrometer during the experiment.
Electronics noise was simulated by Gaussian smearing of reconstructed deposited energies
with measured RMS values.

\begin{figure}[!ht]
\begin{center}
\includegraphics[bb=3cm 0cm 20cm 26cm, scale=0.35, angle=270]{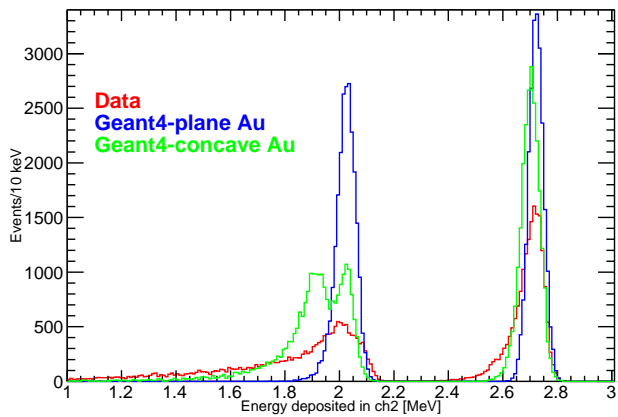}
\caption{\label{fig:edep_i}Energy deposited in single diamond sensor of the spectrometer
under thermal neutron irradiation in comparison with Geant4 simulations
normalized to the measured neutron fluence.
Blue histogram represents Geant4 simulations with plane Au contacts,
while green histogram shows same for concave shape contacts.}
\end{center}
\end{figure}

The scale of energy deposited in both sensors was calibrated by using
digitizer baseline data and $t$-peak position, as the highest and narrowest structure in the spectra.
$t$-peak position was corrected for the energy loss in LiF layer, Air and Au contacts
using Geant4 Monte Carlo.
The comparison of the obtained deposited energy distribution in
single diamond sensor with Geant4 simulation is shown in Fig.~\ref{fig:edep_i}.

The peaks at 2.7 MeV and 2 MeV are due to the absorption of $t$ and $\alpha$, respectively.
Because the energy loss of $t$ is much smaller than that of the $\alpha$ (up to factor of ten for Au contacts)
$t$ peak is higher and narrower. RMS of $t$ peak was about 35 keV,
made of 24 keV due to the first amplifier intrinsic noise
and 25 keV related to energy loss fluctuations
and non-uniformity of contacts. 
$\alpha$ peak in the data has very asymmetric shape related to large energy losses.
The gold metallization film had significant non-uniformity resulting in
variations of the energy loss across sensor active area.
In particular, at the borders of LiF film the underlying contacts were
few times thicker than at the center. This effect was amplified by the choice
of fairly broad LiF coverage aimed to increase detector efficiency.
Geant4 simulations with concave shaped contacts allowed to reproduce
the difference in height between $t$ and $\alpha$ peaks, although
the real shape of the contacts was clearly more complex.

\begin{figure}[!h]
\begin{center}
\includegraphics[bb=3cm 0cm 20cm 26cm, scale=0.35, angle=270]{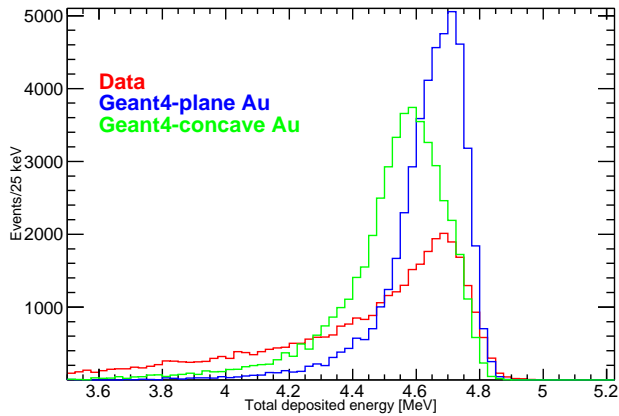}
\caption{\label{fig:triga_edep_tot}Total energy deposited in the spectrometer
by detection of thermal neutrons in comparison with Geant4 simulations
normalized to the measured neutron fluence.
Blue histogram represents Geant4 simulations with plane Au contacts,
while green histogram shows same for concave shape contacts.}
\end{center}
\end{figure}

For thermal neutrons the total energy deposited in the spectrometer corresponds to
$n(^6Li,t)\alpha$ reaction $Q$-value (4.7 MeV).
The comparison of the measured total deposited energy with Geant4 simulations
is shown in Fig.~\ref{fig:triga_edep_tot}.
Also in this spectrum the measured peak exhibits a large tail at its l.h.s.
due to energy loss. Assuming that r.h.s. shoulder of the peak is not
altered by the energy loss we obtained spectrometer total energy resolution of $72$ keV (RMS),
similar to that found in Ref.~\cite{sdw_calib} with charge sensitive amplifiers.

%

Using data from the calibrated fission chamber, located at about 1 cm distance,
we obtained the absolute efficiency of the spectrometer to thermal neutrons of $2.3 \times 10^{-5}$ 1/nv.
This value corresponds to about 55\% of interaction probability in 100 nm of $^6$LiF.
About 4.5\% out of remaining inefficiency is related to the restricted angular acceptance for produced
$t$+$\alpha$ pairs escaping through 300 $\mu$m air gap between the crystals.
Among the remaining 40\% inefficiency, 25\% is due to the aforementioned loss of $\alpha$ or $t$
in non-sensitive parts of the spectrometer, while the other 15\% is related either
to the loss of both reaction products or to the uncertainty on the LiF thickness.

\begin{figure}[!h]
\begin{center}
\includegraphics[bb=3.5cm 2cm 20cm 26cm, scale=0.35, angle=270]{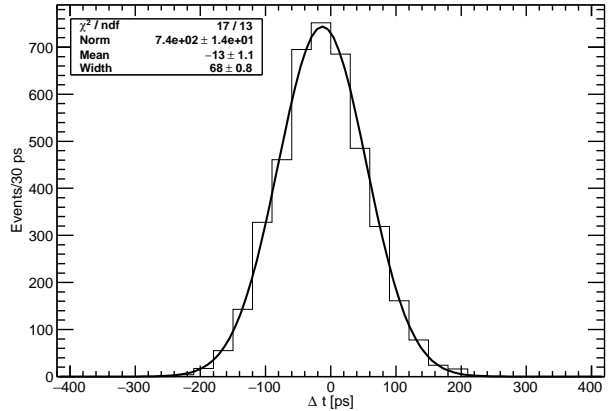}
\caption{\label{fig:triga_dt_peaks}Coincidence time between the two diamonds of the spectrometer
for $\alpha$-$t$ pairs nearly orthogonal to detector plane.}
\end{center}
\end{figure}

The spectrometer demonstrated a very good timing resolution. In fact, the coincidence
time difference between two diamonds shown in Fig.~\ref{fig:triga_dt_peaks} for nearly normally incident $\alpha$-$t$ pairs
had RMS of 68 ps. This corresponds to a single diamond FWHM resolution of 270 ps/($E_{dep.}$/1 MeV),
which is only 30\% worse than the best resolution found in Ref.~\cite{amp_preprint}
with amplifiers directly connected to the diamond detectors.
But it is an order of magnitude better than in Ref.~\cite{sdw_dt} with similar setup
and 40\% better than in Ref.~\cite{nn_fng} for coincidences of two consecutive elastic $n$$-$$C$ scatterings.
Larger PCB thickness and intermediate decoupling ground plane could improve
timing resolution further, but it would reduce detection efficiency
due to lower acceptance for $\alpha$ and $t$ and degradation of energy resolution
due to larger energy loss in air.

The systematic uncertainty of the absolute detection efficiency was dominated
by fission chamber counting rate-to-flux calibration factor, and it was estimated to be 8\%~\cite{altieri_fc}.

\section{Measurement at DD neutron generator}\label{sec:fng}
The spectrometer response to quasi-monochromatic 2.5 MeV neutrons
was measured at FNG facility of ENEA~\cite{fng_facility}.
The detector was installed at 90 degrees with respect to the deuteron beam direction
at distance of 2.8 cm from TiD target center.
The spectrometer was positioned in such a way that the direct neutron flux
from the target impacted normally on diamond surfaces.
The first sensor seen from the target (channel 2) had LiF converter,
while more distant sensor (channel 1) had no converter.
Neutron yield on target was measured by the recoil detector installed inside the beamline
and, on average, it was around $3.4\times 10^8$ n/s. At the detector location
this corresponded to the neutron flux of $2.4\times 10^6$ n/cm$^2$s.

Geant4 simulations of this experiment were made using a realistic neutron spectrum calculated by MCNP simulations~\cite{fng_mcnp}
in the spectrometer location. Except for 2.5 MeV DD-neutrons this spectrum included also
0.25\% contamination of 14.1 MeV DT-neutrons
as well as a tail of scattered neutrons at lower energies. An additional moderator
installed at 50 cm distance at 0 degrees for on-line calibration and detector aluminum support were not
included in MCNP description of neutron source. Therefore, the simulation underestimates
thermal neutron contribution and does not include neutrons scattered from the detector support.

Diamond sensor thresholds were set to 1 MeV as for the measurement in TRIGA.
The coincidence trigger was used over most of run time, few data were taken with
uncorrelated trigger.

The energy calibration was performed using the position of $^9$Be$+\alpha$ peak
due to 14.1 MeV neutron reaction on $^{12}$C. This peak was seen also in coincidence
when $\alpha$, produced in the first diamond, reached the second crystal
with sufficient energy.

\begin{figure}[!h]
\begin{center}
\includegraphics[bb=4.5cm 0cm 20cm 26cm, scale=0.35, angle=270]{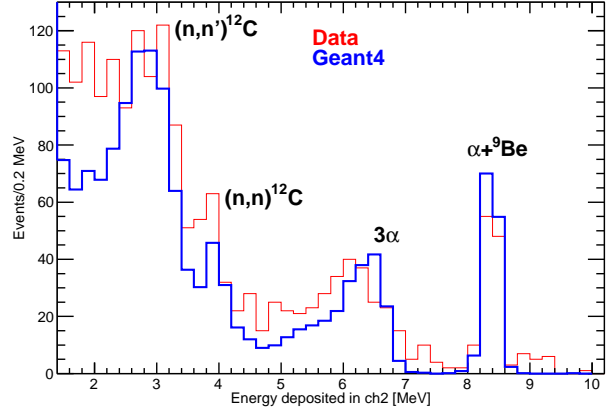}
\caption{\label{fig:fng_edep_or}Energy deposited in single diamond of the spectrometer
in comparison with Geant4 simulations normalized to the measured neutron fluence.
The peak at 8.4 MeV, due to $(n,\alpha)^{12}$C reaction,
has been used for energy calibrations, broad asymmetric peak at 6.5 MeV
is dominated by 3$\alpha$ break-up of $^{12}$C,
peak at 4 MeV is the head of elastic $(n,n)$ scattering
and the one at 2.8 MeV is due to $(n,n^\prime)$ reaction.}
\end{center}
\end{figure}

The energy deposited in a single sensor of the spectrometer, measured with uncorrelated
trigger is shown in Fig.~\ref{fig:fng_edep_or}.
This distribution is dominated by the scattering of 14.1 MeV neutrons on $^{12}$C
of the diamond sensor. Despite the small size of DT contamination in the overall neutron
yield it represents the main contribution in uncorrelated spectra. This is because
spectrometer active volume is made of 2$\times$300 $\mu$m of $^{12}$C
and only 100 nm of LiF converter. Hence, the rate of scattering off $^{12}$C is
enhanced by three orders of magnitude with respect to reactions on $^6$Li.
Furthermore, 2.5 MeV neutrons have not enough energy to induce inelastic nuclear reactions on $^{12}$C
and the only allowed channel for them is the elastic scattering, whose maximum deposited energy (0.7 MeV)
is below threshold.
Instead, 14.1 MeV neutrons may induce a number of inelastic reactions on $^{12}$C,
and even elastic scattering produces recoils with energy ($<$4 MeV) above the threshold.
Thus, from the comparison with Geant4 simulations, also shown in Fig.~\ref{fig:fng_edep_or},
one can observe a number of structures due to the scattering of 14.1 MeV neutrons.
In particular, the peak at 8.4 MeV is due to $(n,\alpha)^{12}$C reaction,
the broad asymmetric peak at 6.5 MeV is dominated by 3$\alpha$ break-up of $^{12}$C,
the peak at 4 MeV corresponds to the head of elastic $(n,n)$ scattering
and the one at 2.8 MeV is due to $(n,n^\prime)$ reaction.
At the energies below 2.5 MeV thermal neutron conversion on $^6$Li adds significant contribution.
Here the rapid increase of $n+^6$Li cross section compensates the difference
in the number of atoms with $^{12}$C of the diamonds. Moreover,
thermal neutrons have a low capture cross section on $^{12}$C
and the energy released in this reaction mostly escapes from the sensor volume.
Underestimation of this region by the simulations indicates that simulated
flux has less thermal neutrons, as expected. But this difference
did not affect the comparison at 2.5 MeV neutron energy.

\begin{figure}[!h]
\begin{center}
\includegraphics[bb=3.5cm 2cm 20cm 26cm, scale=0.35, angle=270]{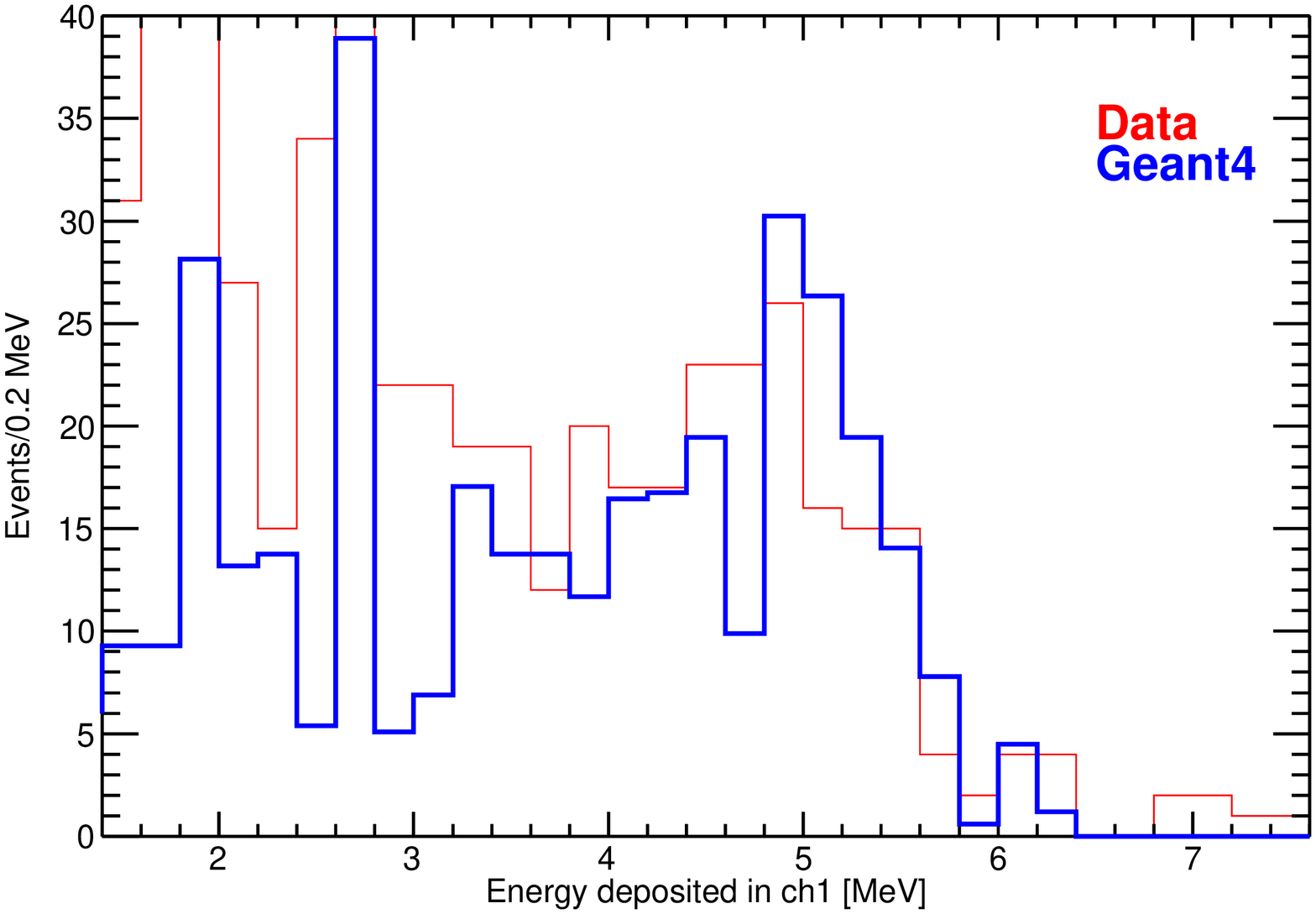}\\
\includegraphics[bb=3.0cm 2cm 20cm 26cm, scale=0.35, angle=270]{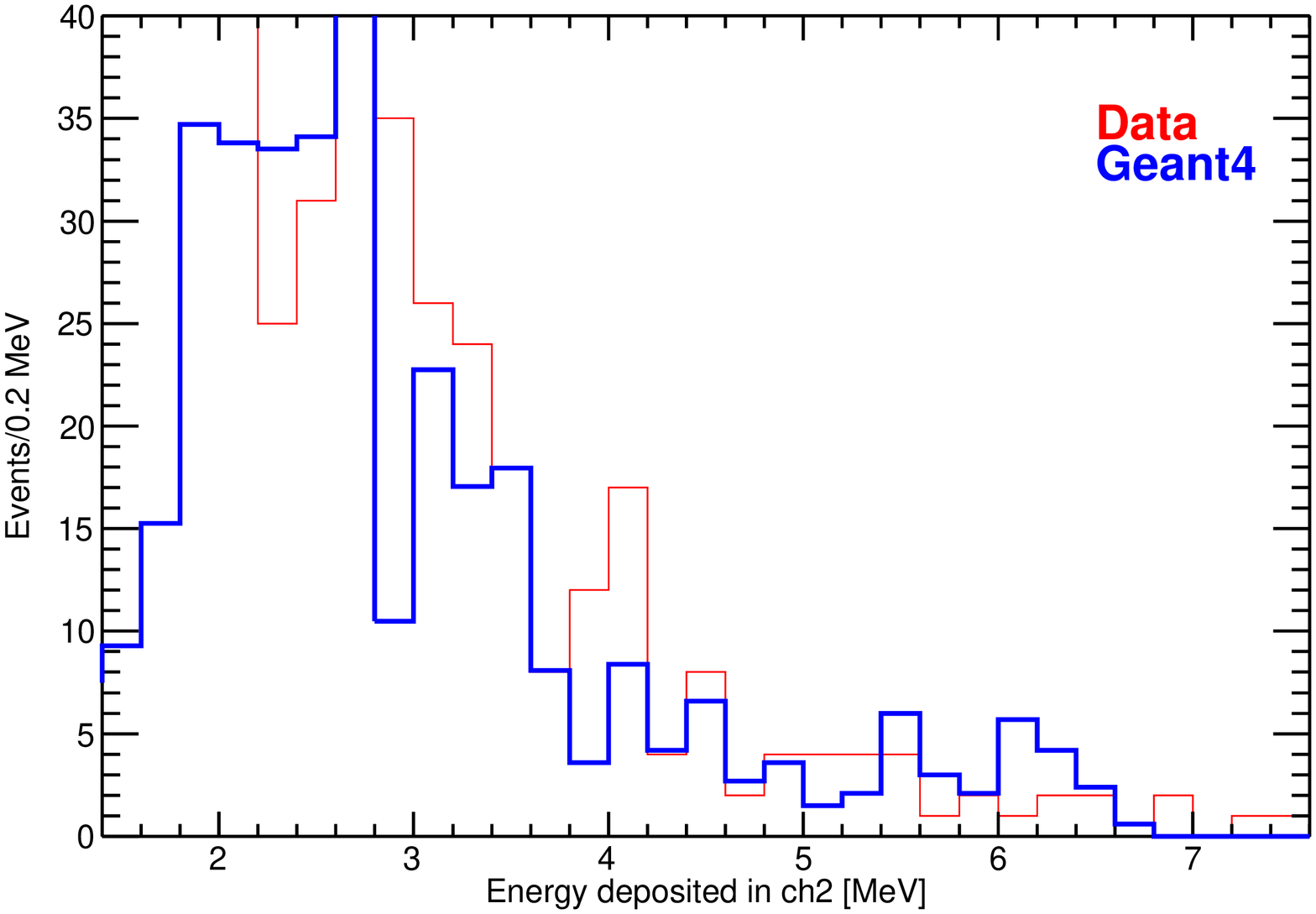}
\caption{\label{fig:fng_edep_coin} The same as if Fig.~\ref{fig:fng_edep_or}, but measured
in coincidence between two diamond detectors.
The neutron flux was normally impacting on ch2 and leaving the spectrometer through ch1.}
\end{center}
\end{figure}

The single sensor deposited energy distributions taken in coincidence, shown in Fig.~\ref{fig:fng_edep_coin},
are very different from those measured with uncorrelated trigger.
Because most of reaction products generated by 14.1 MeV neutrons on $^{12}$C stop
in the bulk of the same diamond the contribution of this part of neutron spectrum
is strongly suppressed. Instead, most of events come from the conversion of
thermal and 2.5 MeV neutrons on $^6$Li. But the conversion of thermal neutrons
is symmetric and produces similar distributions below 3 MeV in both diamond sensors,
while 2.5 MeV neutron conversion features significant Lorentz boost
enhancing energy deposition in the downstream sensor.

The resolution of the detector in this experiment was similar to that in the TRIGA
measurement as one can see from the thermal neutron induced $t$-peak (at 2.7 MeV) RMS of 34 keV.
The $^9$Be$+\alpha$ peak showed RMS of 100 keV, dominated by the incident neutron energy uncertainty.

\begin{figure}[!ht]
\begin{center}
\includegraphics[bb=4.5cm 0cm 20cm 26cm, scale=0.35, angle=270]{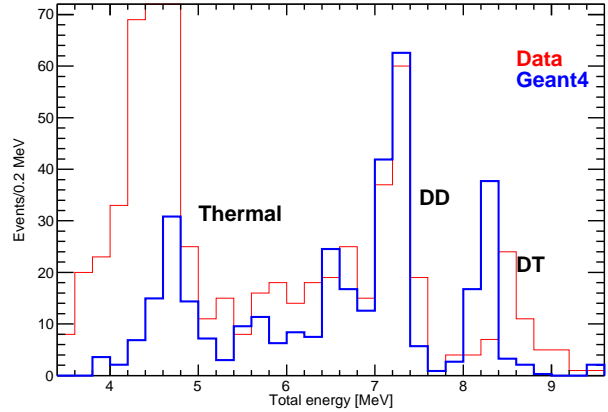}
\caption{\label{fig:etot_dd}Total energy deposited in the spectrometer
measured in coincidence at FNG with TiD target
in comparison with Geant4 simulations normalized to the measured neutron fluence.}
\end{center}
\end{figure}

The total deposited energy distribution shown in Fig.~\ref{fig:etot_dd} exhibits
three peaks:
peak at 8.4 MeV is due to remaining $^9$Be$+\alpha$ produced by 14.1 MeV neutrons
at the surface of the first diamond,
peaks at 7.2 MeV and 4.7 MeV are due to conversion on $^6$Li of 2.5 MeV and thermal neutrons, respectively.
The data are in good agreement with Geant4 simulations except for underestimated thermal neutron contribution.

In the measured data the peak widths are enhanced by the spectrometer energy resolution RMS
up to 83 keV for thermal neutron peak and 98 keV for 2.5 MeV neutrons.
The last value includes also incident neutron energy uncertainty.

%

The total number of events in DD-peak was about 130, which allowed to confirm
the expected spectrometer absolute efficiency value of $4.5\times 10^{-9}$ 1/nv at neutron energy of 2.5 MeV with
9\% statistical and 20\% systematic~\cite{sdw_calib} uncertainties.

\section{Conclusions}\label{sec:conclusions}
New prototype of compact neutron spectrometer for high flux environments
based on $^6$Li converter and diamond detectors was assembled.
This included a number of improvements aimed to resolve various issues observed in Refs.~\cite{sdw_calib,sdw_tapiro,sdw_dt}.
In particular, the new spectrometer was built with commercial single crystal CVD diamonds of electronic grade.
These crystals feature a better charge collection as well as a higher radiation hardness
as shown in Refs.~\cite{sdw_dt,pillon}.
Ohmic metal contacts were deposited on the diamonds, suppressing build-up of space
charge observed in the previous prototypes~\cite{sdw_tapiro}.
In the new prototype no space charge effects were observed.
New passive preamplification of signal at detector side was implemented to improve its energy resolution.
This preamplification is based on RF transformer, not sensitive to high neutron flux.
Compact mechanical design was implemented and allowed to reduce detector size to a 13 cm long tube
with diameter of 1 cm.

The spectrometer was tested in the thermal column of
TRIGA reactor and at DD neutron generator.
It demonstrated good performances as energy resolution of 72 keV (RMS)
and coincidence timing resolution of 68 ps (RMS).
The measured data are overall in agreement with Geant4 simulations.
The only remaining difference is related to excessive energy loss
of $\alpha$s produced by neutrons on $^6$Li. This effect may be mitigated
in future prototypes by improving contact uniformity
and reducing the area of LiF converter.

\section*{Acknowledgement}
Authors would like to acknowledge the excellent support provided during the experiments
by the staff and technical services of LENA and FNG facilities.
We thank Prof.~Maurizio Canepa of Genova University for the characterization of deposited LiF
films by ellipsometry technique.
This work was supported by the Istituto Nazionale di Fisica Nucleare INFN-E project.

\bibliographystyle{elsarticle-num}
\bibliography{sdw_single_pcb}

\begin{thebibliography}{10}
\expandafter\ifx\csname url\endcsname\relax
  \def\url#1{\texttt{#1}}\fi
\expandafter\ifx\csname urlprefix\endcsname\relax\def\urlprefix{URL }\fi
\expandafter\ifx\csname href\endcsname\relax
  \def\href#1#2{#2} \def\path#1{#1}\fi

\bibitem{cvd_rad_hard}
W.~de~Boer, et~al., Phys. Status Solidi 204 (2007) 3009.

\bibitem{sdw_calib}
M.~Osipenko, et~al., Nucl. Instr. and Meth. A 799 (2015) 207.

\bibitem{sdw_tapiro}
M.~Osipenko, et~al., Test of a prototype neutron spectrometer based on diamond
  detectors in a fast reactor, in: Proceedings of 4th International Conference
  on Advancements in Nuclear Instrumentation Measurement Methods and their
  Applications (ANIMMA 2015), IEEE Nucl.Sci.Symp.Conf.Rec., Lisbon, Portugal,
  2015.
\newblock \href {http://dx.doi.org/10.1109/ANIMMA.2015.7465605}
  {\path{doi:10.1109/ANIMMA.2015.7465605}}.

\bibitem{sdw_dt}
M.~Osipenko, et~al., Nucl. Instr. and Meth. A 817 (2016) 19.

\bibitem{cvd_e6}
\href{http://www.e6cvd.com}{{Element Six, Electronic grade CVD diamonds}}.
\newline\urlprefix\url{http://www.e6cvd.com}

\bibitem{Trucchi1}
D.~Trucchi, et~al., IEEE Electron Device Lett. 33 (2012) 615.

\bibitem{Trucchi2}
M.~Girolami, et~al., Phys. Status Solidi A 212 (2015) 2424.

\bibitem{Trucchi3}
M.~Rebai, et~al., Journal of Instrumentation 10.

\bibitem{Trucchi4}
C.~Cazzaniga, et~al., Nucl. Instr. and Meth. B 405 (2017) 1.

\bibitem{esolder}
\href{http://epoxy-produkte.de}{{EPOXY Produkte GmbH}}.
\newline\urlprefix\url{http://epoxy-produkte.de}

\bibitem{Mini-Circuit}
\href{http://www.minicircuits.com}{{Mini-Circuits}}.
\newline\urlprefix\url{http://www.minicircuits.com}

\bibitem{SCK-CEN}
\href{http://science.sckcen.be/en/Facilities/VENUS}{{VENUS-F reactor (SCK-CEN,
  Mol Belgium)}}.
\newline\urlprefix\url{http://science.sckcen.be/en/Facilities/VENUS}

\bibitem{wantcom}
\href{http://www.wantcominc.com}{{WanTcom, Inc.}}
\newline\urlprefix\url{http://www.wantcominc.com}

\bibitem{lena_pv}
M.~Prata, et~al., Eur. Phys. J. Plus 129 (2014) 255.

\bibitem{altieri_fc}
A.~Altieri, Private communication.

\bibitem{triga_mon}
A.~{Borio~di~Tigliole}, et~al., Home-made refurbishment of the instrumentation
  and control system of the triga reactor of the university of pavia, 2008.

\bibitem{triga_flux}
N.~Protti, S.~Bortolussi, M.~Prata, P.~Bruschi, S.~Altieri, D.~Nigg, Neutron
  spectrometry for the university of pavia triga thermal neutron source
  facility, Vol. 107, TRANSACTIONS OF THE AMERICAN NUCLEAR SOCIETY, 2012, p.
  1269.

\bibitem{geant4}
S.~Agostinelli, et~al., Nucl. Instr. and Meth. 506 (2003) 250.

\bibitem{amp_preprint}
M.~Osipenko, et~al., {INFN} preprint, {INFN-13-17-GE}, 2014.

\bibitem{nn_fng}
M.~Osipenko, et~al., Eur. Phys. J. Plus 129 (2014) 268.

\bibitem{fng_facility}
M.~Martone, M.~Angelone, M.~Pillon, J. of Nucl. Mater. B 212 (1994) 1661.

\bibitem{fng_mcnp}
M.~Angelone, M.~Pillon, P.~Batistoni, M.~Martini, M.~Martone, V.~Rado, Rev. of
  Sci. Instrum. 67 (1996) 2189.

\bibitem{pillon}
M.~Pillon, et~al., Jour. of Appl. Phys. 104 (2008) 054513.

\end{thebibliography}

\end{document}